\documentclass[twocolumn]{aastex631}
\let\tablenum\relax
\usepackage{amsfonts,amsmath,graphicx,natbib,url,hyperref,nicefrac,siunitx}
\usepackage{amssymb,verbatim,subfigure,paralist,soul,comment}

\hypersetup{colorlinks=true, urlcolor=blue, citecolor=blue}

\usepackage{color}

\begin{document}

\title{Quantifying Advantages of a Moving Mesh in Nuclear Hydrodynamics}

\author{Dillon L. Hasenour}
\affiliation{Department of Physics and Astronomy, Purdue University, 525 Northwestern Avenue, West Lafayette, IN 47907, USA}

\author{Paul C. Duffell}
\affiliation{Department of Physics and Astronomy, Purdue University, 525 Northwestern Avenue, West Lafayette, IN 47907, USA}

\begin{abstract}

Many astrophysical explosions, such as type Ia supernovae, classical novae, and X-ray bursts, are dominated by thermonuclear runaway. To model these processes accurately, one must evolve nuclear reactions concurrently with hydrodynamics. We present an application of the moving mesh technique to this field of computation with the aim of explicitly testing the advantages of the method against the fixed mesh case. By way of traditional Strang splitting, our work couples a 13 isotope nuclear reaction network to a 1D moving mesh, Cartesian geometry hydrodynamics code. We explore three reacting problems: an acoustic pulse, a burning shock, and an advecting deflagration. Additionally using the shock jump conditions, we semi-analytically solve the burning shock problem under the assumption of quick, complete burning with the hope of establishing a useful and easy to set-up test problem. Strong moving mesh advantages are found in advecting, deflagrating flame fronts, where the technique dramatically reduces numerical diffusion that would otherwise lead to very fast artificial deflagration.  

\end{abstract}


\section{Introduction} \label{sec:intro}

High energy astrophysical events often involve the process of nuclear burning, and in some cases, thermonuclear runaway is the dominant mechanism of interest. This is true, for example, in type Ia supernovae \citep{Pakmor_2013, Shen_2018, Polin_2019}, classical novae \citep{Starrfield_2020}, and X-ray bursts \citep{Eiden_2020}. To hydrodynamically model these events, it becomes necessary to include a coupled nuclear network that accounts for the energy generation from burning, which in turn introduces a myriad of difficulties and subtleties to conventional codes. The core challenge is to evolve the stiff ODE behavior of reaction networks from the steep temperature dependence of reaction rates at hydrodynamical timescales.

Much progress has been made by Eulerian finite volume methods utilizing Adaptive Mesh Refinement (AMR) techniques. \texttt{FLASH} \citep{Fryxell_2000} and \texttt{CASTRO} \citep{Almgren_2010} are two leading examples of AMR codes for nuclear hydrodynamics. AMR refines areas of interest (such as where burning occurs) to improve spatial and temporal resolution, while evolving the less computationally demanding areas at lower resolutions. Additionally, much work has gone into developing reaction networks suitable for astrophysical flows, including the \texttt{iso7}/\texttt{aprox13}/\texttt{aprox21}/\dots suite \footnote{\texttt{aprox13} and more are available at \url{https://cococubed.com/}} \citep{Timmes_1999} and recently \texttt{Pynucastro} \citep{Smith_2023}. Networks with more isotopes and rates generally are better models as more physics is included. However, they are computationally much more expensive, and smaller networks, if chosen carefully, can often capture the energy generation well enough \citep{Timmes_2000c}.  As always, the optimal balance depends on the astrophysical problem being solved.

As used by the code \texttt{AREPO} (\cite{Springel_2010} for hydrodynamical description and \cite{Pakmor_2013} for early work with nuclear networks), another approach is to use the moving mesh technique, which broadly refers to a class of finite volume hydrodynamic models with meshes that are allowed to move according to local fluid velocities. This is open to many choices of mesh structure varying in geometries and mesh motion. For example \texttt{AREPO}, \texttt{TESS} \citep{Duffell_2011}, and \texttt{RICH} \citep{Yalinewich_2015}, have unstructured tessellations, and \texttt{DISCO} \citep{Duffell_2016}, \texttt{JET} \citep{Duffell_2018}, and \texttt{SPROUT} \citep{Mandal_2023}, have unique structured meshes designed for specific problems. In general, the moving mesh technique has been shown to dramatically reduce the effects of numerical diffusion \citep{Yalinewich_2015} and permit much longer timesteps for fluids experiencing bulk motion. 

Extending the hydrodynamical advantages of moving mesh to include a reaction network, the reduction of numerical diffusion logically extends to minimizing the artificial mixing of fuel and ash in problems with burning. However, it has not been explicitly shown in test problems to be significant. We have developed a prototype 1D nuclear hydrodynamics code on a moving mesh in order to quantify these advantages by close comparisons to the fixed mesh case.

This work proceeds as follows. We describe our specific numerical methodology behind moving mesh hydrodynamics, nuclear networks, and coupling in $\S$\ref{sec:method}. In $\S$\ref{sec:detonation}, we introduce the semi-analytic solution to a one-dimensional test problem in the limit of quick burning that is useful in understanding a latter application. We present the performance of the testbed code across three problems in $\S$\ref{sec:code} including demonstrations of overall convergence and moving mesh improvements. Finally we summarize and discuss in $\S$\ref{sec:discussion}.

\section{Numerical Methods} \label{sec:method}

The testbed code used in this work is a one-dimensional, finite volume, Cartesian geometry, moving mesh hydrodynamics $+$ nuclear reactions code. In the following sections, we will describe the methods and techniques used to evolve hydrodynamics, to evolve reactions, and to couple them together. 

\subsection{Strang Splitting}
To couple hydrodynamics with nuclear reactions, we are using a traditional Strang split approach \citep{Strang_1968}. The overall numerical scheme can be organized into a set of conservation laws in mass, momentum, and energy as

\begin{equation}
    \label{eq:Operator_splitting}
    \frac{\partial \mathbf{U}}{\partial t} = \mathcal{A}(\mathbf{U}) + \mathcal{R}(\mathbf{U}),
\end{equation}
where $\mathbf{U}$ is the vector of conserved quantities, $\mathcal{A}(\mathbf{U})$ are terms due to hydrodynamics, and $\mathcal{R}(\mathbf{U})$ are terms due to reactions. The operator splitting approach progresses $\mathbf{U}$ by alternating between evolving according to $\mathcal{R}$ omitting $\mathcal{A}$ and evolving according to $\mathcal{A}$ omitting $\mathcal{R}$. Strang splitting is a second order coupling method that improves upon first order operator splitting. For a given timestep, we evolve the conserved quantities for half the total time according to reactions, then evolve the result for the full time according to hydrodynamics, and finally evolve that result according to reactions for the remaining half total time. This scheme effectively centers the reactions in time over one timestep. Symbolically,

\begin{equation}
    \label{eq:Strang_splitting}
    \begin{gathered}
        \left[\frac{\partial \mathbf{U}^n}{\partial t} = \mathcal{R}(\mathbf{U}^n)\right]_{\Delta t/2} \longrightarrow \mathbf{U}^* \\
        \left[\frac{\partial \mathbf{U}^*}{\partial t} = \mathcal{A}(\mathbf{U}^*)\right]_{\Delta t} \longrightarrow \mathbf{U}^{**} \\
        \left[\frac{\partial \mathbf{U}^{**}}{\partial t} = \mathcal{R}(\mathbf{U}^{**})\right]_{\Delta t/2} \longrightarrow \mathbf{U}^{n+1},
    \end{gathered}
\end{equation}
where $\mathbf{U}^n$ is the state at timestep $n$, $\mathbf{U}^*$ and $\mathbf{U}^{**}$ are intermediate states, and $\Delta t$ is the length of the timestep. 

\subsection{Hydrodynamics}
When evolving according to $\mathcal{A}(\mathbf{U})$, the testbed code solves 1D Euler's equations in conservation law form:

\begin{equation}
\label{eq:convo_laws}
    \begin{gathered}
        \partial_t(\rho) + \partial_x( \rho v ) = 0 \\
        \partial_t( \rho v ) + \partial_x ( \rho v^2 + P ) = 0 \\
        \partial_t\left( \frac{1}{2}\rho v^2 + \epsilon \right) + \partial_x \left( \left[ \frac{1}{2}\rho v^2 + \epsilon + P \right]v \right) = 0 \\
        \partial_t(\rho X_i) + \partial_x( \rho v X_i ) = 0,
    \end{gathered}
\end{equation}
where $\rho$, $v$, $P$, $\epsilon$, and $X_i$ are the primitive variables and stand for density, velocity, pressure, internal energy density, and mass fraction of isotope $i$ respectively. As reactions are held fixed during this update, the masses of individual isotopes are also conserved, and the mass fractions are treated like passive scalars. In general, the right hand side can contain source terms such as external gravity, spherical geometry, or nuclear energy generation; however, we instead include the total binding energy of our composition into the total internal energy by considering

\begin{equation}
\label{eq:e_int}
    \epsilon = \epsilon^{(eos)}_{th} (\rho,P,\mathbf{X}) + \sum_i \rho E_{b,i} X_i,
\end{equation}
where $\epsilon^{(eos)}_{th}$ is the thermal energy density from the equation of state, and $E_{b,i}$ is the specific internal energy of isotope $i$. When the composition changes during the reaction network updates, energy is effectively transferred between the binding energy and thermal energy. We use the stellar equation of state \texttt{helmeos} to determine $\epsilon^{(eos)}_{th}$ and to relate the thermodynamic properties when needed elsewhere \citep{Timmes_2000}. In \texttt{helmeos}, the fluid is composed of ideal gas ions, degenerate/relativistic electrons, and radiation, all in local thermodynamic equilibrium (LTE).

Following the finite volume prescription, we discretize the computational domain into a number of zones arranged in a 1D Cartesian mesh. We advance the domain using the integral form of the field equations. At the beginning of each timestep, we make the choice for the interfaces of neighboring zones to move according to the local fluid velocity. In 1D, this is simply the average of the velocities of the zones to the left and right of the interface. This necessitates a correction to the fluxes, i.e. we must subtract off the conserved material overtaken by the interface from the flux. The hydrodynamical update for conserved quantities of mass, momentum, energy, and individual mass of isotope follows the form

\begin{equation}
\label{eq:timestep}
    \begin{gathered}
    M^{n+1}_i = M^{n}_i - \Delta t\big[\Delta A (F-wU)\big]_{i+1/2} \\
     + \Delta t\big[\Delta A (F-wU)\big]_{i-1/2} ,
    \end{gathered}
\end{equation}

where $M^{n}_i$ is the conserved quantity in zone $i$ at timestep $n$, $\Delta t$ is the length of the timestep, and the left and right interface quantities of flux $F$, velocity $w$, and conserved quantity $U$ are denoted by $i\pm 1/2$. In 1D Cartesian geometry, we can set the surface area of the interfaces $\Delta A$ to be uniformly unit. The flux across the interface is approximated via the HLLC Riemann solver \citep{Toro_1994}. The HLLC solver accounts for the existence of a contact discontinuity and pairs well with the moving mesh technique to lower numerical diffusion, as long as we correctly choose the solution along the path traced out by the interface \citep{Duffell_2011}.

To achieve second order convergence, we use the Piecewise Linear Method (PLM) and a method of lines Runge-Kutta scheme (RK). PLM interpolates zone centered values to interface centered values in order to achieve higher order spatial accuracy. It utilizes a \textit{minmod} slope limiter to avoid spurious oscillations and ensure stability. 
Since we are explicitly evolving all isotopes, PLM over mass fractions does not preserve $\sum_i  X_i = 1$. Consequently, we renormalize the PLM interpolated mass fractions before handing the result to the Riemann solver. Regarding higher order accuracy in time, we use the method of lines RK2 scheme described in \cite{Shu_1988}. See Section \ref{sec:ap} for a demonstration of the testbed code's achieved overall convergence.

\subsection{Reactions}
A nuclear reaction network is a collection of rates that describe how a composition of elements/isotopes evolve in time. The network used in this work is \texttt{aprox13} \citep{Timmes_1999}. It contains 13 explicit isotopes $^4$He, $^{12}$C, $^{16}$O, $^{20}$Ne, $^{24}$Mg, $^{28}$Si, $^{32}$S, $^{36}$Ar, $^{40}$Ca, $^{44}$Ti, $^{48}$Cr, $^{52}$Fe, and $^{56}$Ni. The processes of key interest are the triple alpha process, the alpha chain, heavy ion interactions, and approximations of $(\alpha,p)(p,\gamma)$ links through intermediate isotopes. 

Traditional Strang splitting evolves the reaction network in time with hydrostatic burning, i.e. the density and temperature are held fixed. Typically, we evolve over molar abundances $Y_i = X_i/A_i$ (mass fraction over mass number). The time derivatives can now be expressed in the ODE formulation of Equation \ref{eq:rxn_ODE}:

\begin{equation}
    \label{eq:rxn_ODE}
    \frac{dY_i}{dt}=f_i(\{Y_i\}),
\end{equation}
where $f_i$ is commonly referred to as the right-hand side (RHS) and contains all the rates affecting isotope $i$. For a given timestep, the testbed code progresses the reaction network using a $4^{th}$ order, semi-implicit, generalized step-doubling method ODE integrator. This integrator is described in detail in Appendix \ref{sec:integrator}. Given a density, temperature, and composition, \texttt{aprox13} computes the RHS and corresponding Jacobian ($\partial f_i/\partial X_j$) as needed. Using the density and total energy from the beginning of the reaction step, the final composition updates the thermal energy and temperature per the equation of state. 

\subsection{Thermal Diffusion}
For the purpose of exploring a deflagration flame front in this work, we can include a thermal diffusion flux in the energy conservation equation. Functionally, this modifies the third row of Equation \ref{eq:convo_laws} to be

\begin{equation}
\label{eq:therm_diff}
    \partial_t\left( \frac{1}{2}\rho v^2 + \epsilon \right) + \partial_x \left( \left[ \frac{1}{2}\rho v^2 + \epsilon + P \right]v \right) = \partial_x \left( k_{th} \nabla T\right),
\end{equation}
where $k_{th}$ is the thermal conductivity at the interface, and $\nabla T$ is the local temperature gradient. In 1D, to first order, the temperature gradient is the slope in temperature between neighboring zones. The thermal conductivities are determined by the stellar opacity code \texttt{sig99}, which takes  density, temperature, and composition of the fluid as an input \citep{Timmes_2000b}. Since we need the conductivity at the interface, we average over the zone centered values by taking

\begin{equation}
\label{eq:therm_con}
    k_{th} = \frac{k_{th}(\rho_L,T_L,\mathbf{X}_L)+k_{th}(\rho_R,T_R,\mathbf{X}_R)}{2}.
\end{equation}

\subsection{Timestep}
The timestep is determined by the Courant-Friedrichs-Lewy (CFL) condition common to numerical solutions of hyperbolic systems. The guiding principle is that information should only travel to directly neighboring zones over one timestep, which leads to the following condition on the length of the timestep, based on the sound speed:

\begin{equation}
\label{eq:cfl_condition}
    \Delta t = C_{\mbox{CFL}}*\min_i \left[\frac{\Delta x_i}{c_s(\rho_i,T_i,\mathbf{X}_i)+|v_i-\bar{w}_i|}\right],
\end{equation}
where $\Delta t$ is the timestep, $\Delta x_i$ is the 1D volume of zone $i$, $c_s$ is the sound speed determined by the equation of state, $v_i$ is the local fluid velocity, and $\bar{w_i}$ is the average of the left and right interface velocities. The subtraction of the average mesh motion is a benefit of the moving mesh technique that allows for longer timesteps to be taken. We choose a $C_{\mbox{CFL}}$ number of $0.2$. 

\section{Semi-analytic Detonation} \label{sec:detonation}

In this section, we introduce a semi-analytic burning shock detonation that aims to be as simple as possible in order to be easily replicated by other codes in the field. It is an application of the Rankine-Hugoniot jump conditions with the assumption of quick burning. Consider an initial cold fluid of constant density, pressure, and composition, but with a jump in velocity. We make the choice of a positive velocity left of the discontinuity pointing towards a stationary fluid on the right. 

\begin{equation}
\label{eq:burn_shock_init}
    \begin{gathered}
        \rho_L = \rho_R = \rho_0 \\
        P_L = P_R = P_0 \\
        \mathbf{X}_L = \mathbf{X}_R = \mathbf{X}_0\\
        v_L = v_0, ~v_R = 0.
    \end{gathered}
\end{equation}

After a bit of time, a heated shocked region will form from the collision of the two fluids. If this hydrodynamically heated region reaches sufficient temperature, nuclear reactions will occur quickly and release binding energy further powering the shocks. We argue from global conservation of momentum and symmetry that we can expect the central shocked region to have velocity of $v_0/2$. 

\begin{figure}[h!]
\label{fig:bs_sketch}
    \centering
    \fig{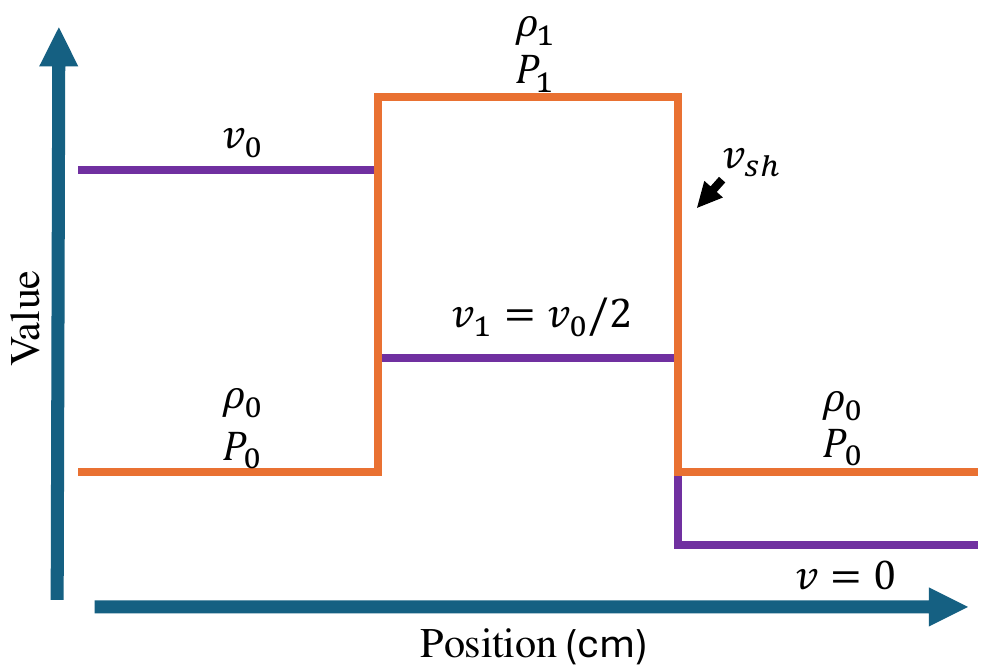}{0.45\textwidth}{}
    \caption{Sketch of burning shock solution. Density and pressure form top-hat structures, while velocity forms a step like structure.}
\end{figure}

Knowing the velocity behind the shock, the final composition, and the initial conditions, we can solve for the shock speed and remaining fluid properties of the shocked region via the shock jump conditions. If we Galilean transform to the frame of the shock, the jump condition merely states that the fluxes across the shock must be equal. Balancing mass, momentum, and energy fluxes results in three equations with three unknowns: the shock velocity, the density behind the shock, and the pressure behind the shock. Equation \ref{eq:burn_shock_jump} sets up the system of equations as

\begin{equation}
\label{eq:burn_shock_jump}
    \begin{gathered}
        \bar{\rho_1} \bar{v_1} = \bar{\rho_0} \bar{v_0} \\
        \bar{\rho_1} \bar{v_1}^2 + \bar{P_1} = \bar{\rho_0} \bar{v_0}^2 + \bar{P_0} \\
        \bar{v_1}\left(\frac{1}{2}\bar{\rho_1}\bar{v_1}^2 + \bar{\epsilon_1} +\bar{P_1}\right) = \bar{v_0}\left(\frac{1}{2}\bar{\rho_0}\bar{v_0}^2 + \bar{\epsilon_0} +\bar{P_0}\right),
    \end{gathered}
\end{equation}
where barred quantities are the Galilean transformed values ($\bar{v}=v-v_{sh}$, while other quantities are unchanged). The unshocked, upstream fluid is denoted with subscript $0$, while the shocked, downstream fluid is subscript $1$. Transforming back into the initial frame and solving in terms of $v_{sh}$, we obtain the following system of equations:

\begin{equation}
\label{eq:burn_shock_sol}
    \begin{gathered}
        \rho_1 = \rho_0 \frac{v_{sh}}{v_{sh}-v_1} \\
        P_1 = P_0 + \rho_0 v_{sh} v_1 \\
        0 = P_0 v_1 + \frac{1}{2} \rho_0 v_{sh} v_1^2 - \rho_0 v_{sh}(\epsilon_1-\epsilon_0).
    \end{gathered}
\end{equation}

Here the internal energies $\epsilon_1$ and $\epsilon_0$ includes both the thermal energy and the binding energy as in Equation \ref{eq:e_int}. If we know the final composition, then the only unknown in the third row of Equation \ref{eq:burn_shock_sol} is $v_{sh}$, because the dependence of $\epsilon_1$ on $\rho_1$ and $P_1$ can be reduced to a dependence on $v_{sh}$ by substitution. Therefore, we can perform a 1D Newton-Rahpson root finding algorithm to solve for $v_{sh}$ for any given set of initial conditions. This gives us a complete picture of the expected solution. 

\section{Code Tests} \label{sec:code}
Now, we present three tests/applications of moving mesh hydrodynamics with nuclear reactions. First, we run an acoustic pulse problem similar to \cite{Zingale_2019} to demonstrate the convergence of our methods with and without reactions. Next, we explore resolution effects and benefits of the natural moving mesh compression of zones in the semi-analytic detonation problem. Finally, we modify the proof-of-concept deflagration flame also in \cite{Zingale_2019} to show that moving mesh gives advantages with respect to numerical diffusion's mixing of fuel and ash. 

\begin{figure*}
\centering
\fig{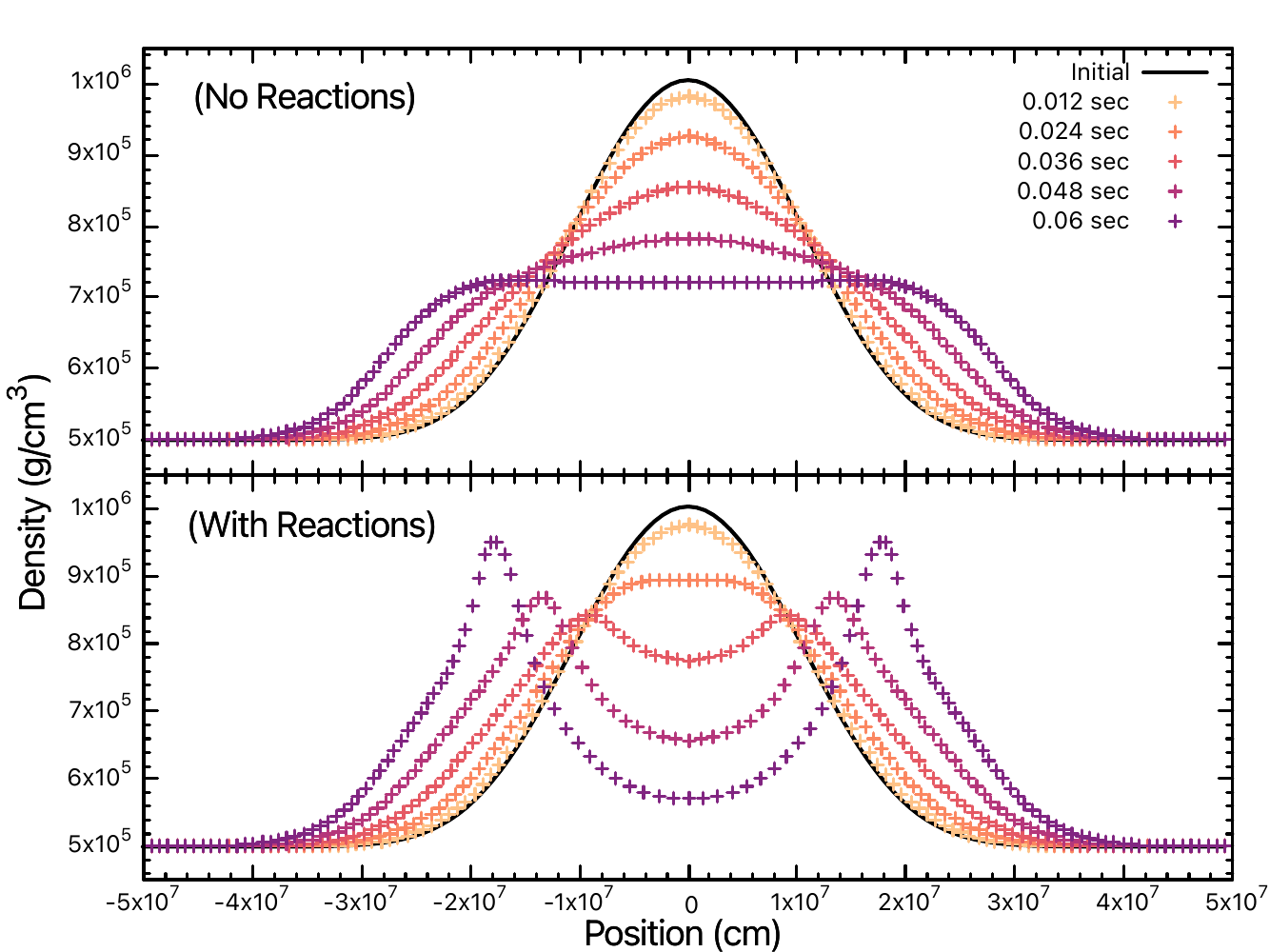}{.75\textwidth}{}
\caption{1D acoustic pulse density vs. position at five checkpoints in time ending when $t=\SI{0.06}{\sec}$. (\textit{Top}) No reactions. Evolves purely hydrodynamically with \texttt{helmeos} equation of state. (\textit{Bottom}) Burning results in a more aggressive expansion and no longer conserves entropy. In both cases, the moving mesh technique is used with 128 computational zones.}
\label{fig:AcousticPulse}
\end{figure*}

\begin{figure}
    \centering
    \includegraphics[width=0.45\textwidth]{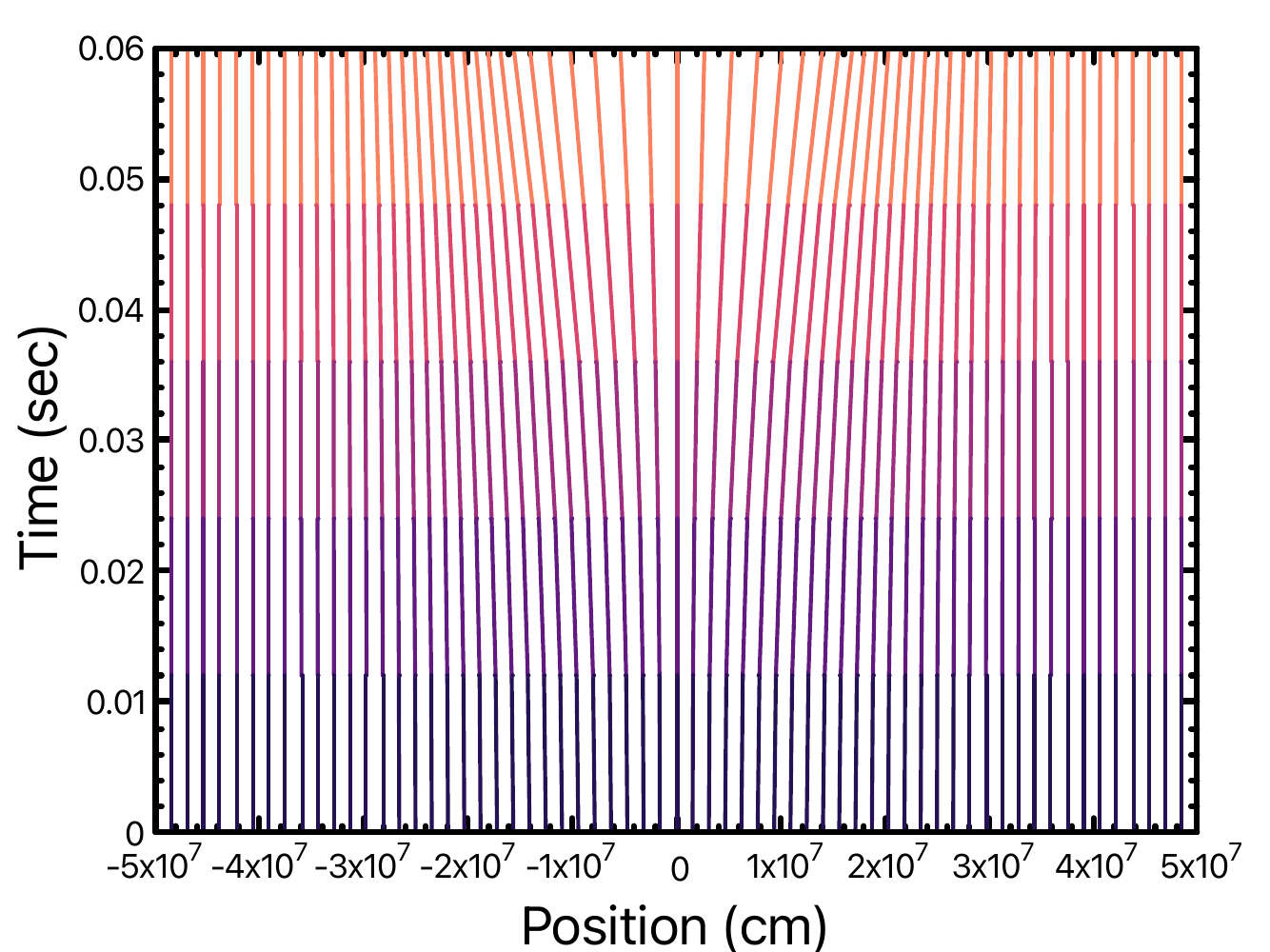}
    \caption{The acoustic pulse mesh motion over time. Each line represents the trajectory of a zone. Only every other zone is plotted.}
    \label{fig:ap_mesh}
\end{figure}

\subsection{Acoustic Pulse} \label{sec:ap}
The acoustic pulse problem is a constant entropy, shockless wave used to demonstrate the overall convergence of our numerical methods. Without burning, entropy will be conserved; however with burning, the release of nuclear binding energy will heat the active region non-adiabatically. The specific setup in this work is a 1D adjustment of the 2D acoustic pulse problem in \cite{Zingale_2019}. We begin by choosing a surrounding density $\rho_0$, temperature $T_0$, and composition $\mathbf{X}_0$. Then, the equation of state finds the surrounding entropy $s_0$ and pressure $P_0$. The initial velocity is zero throughout, and the pressure across the domain follows the smooth distribution

\begin{equation}
\label{eq:init_def_pre}
    P(x) = P_0 \left(1+f_p e^{-\frac{x^2}{\Delta_r^2}} \cos^6\left(\frac{\pi x}{L_0}\right)\right),
\end{equation}
where $x$ is the position, and $f_p$, $\Delta_r$, and $L_0$ are parameters that control amplitude, width, and domain length respectively. Finally, using this pressure and entropy, we use the equation of state to solve for the remaining thermodynamic quantities of interest. The initial composition is pure $^4$He. Figure \ref{fig:AcousticPulse} shows a typical density profile with and without reactions for the parameters given in Table \ref{table:acoustic_pulse}.

\begin{table}[h!]
\centering
\begin{tabular}{||c c||}
 \hline
 Parameter & Value \\ [0.5ex] 
 \hline\hline
 $\rho_0$ & \SI{5e5}{\gram/\cm^3} \\ 
 $T_0$ & \SI{3e8}{\kelvin} \\
 $X_0(\mbox{He})$ & 1.0 \\
 $v_0$ & \SI{0}{\cm/\sec} \\
 \hline
 $f_p$ & 2.0 \\
 $\Delta_r$ & \SI{2e7}{\cm} \\
 $L_0$ & \SI{e8}{\cm} \\ 
 $t$ & \SI{0.06}{\sec} \\
 \hline
\end{tabular}
\caption{(\textit{Top}) Initial hydrodynamic parameters of the acoustic pulse problem. (\textit{Bottom}) Domain parameters.}
\label{table:acoustic_pulse}
\end{table}

Further, Figure \ref{fig:ap_mesh} shows the motion of the mesh over time. Following the local fluid velocity, the mesh expands in the central region and compresses just ahead of the density peaks. The minimum zone volume in the reacting case is a factor $\sim 0.71\times$ the initial volume. Although the moving mesh reaches smaller zone sizes, this effect shortening the timestep is offset by the subtraction of the average mesh motion in Equation \ref{eq:cfl_condition}. The net result is that the total CPU time for the moving mesh is 1432 seconds and 1691 seconds for the fixed mesh.

\begin{figure}
\centering
\fig{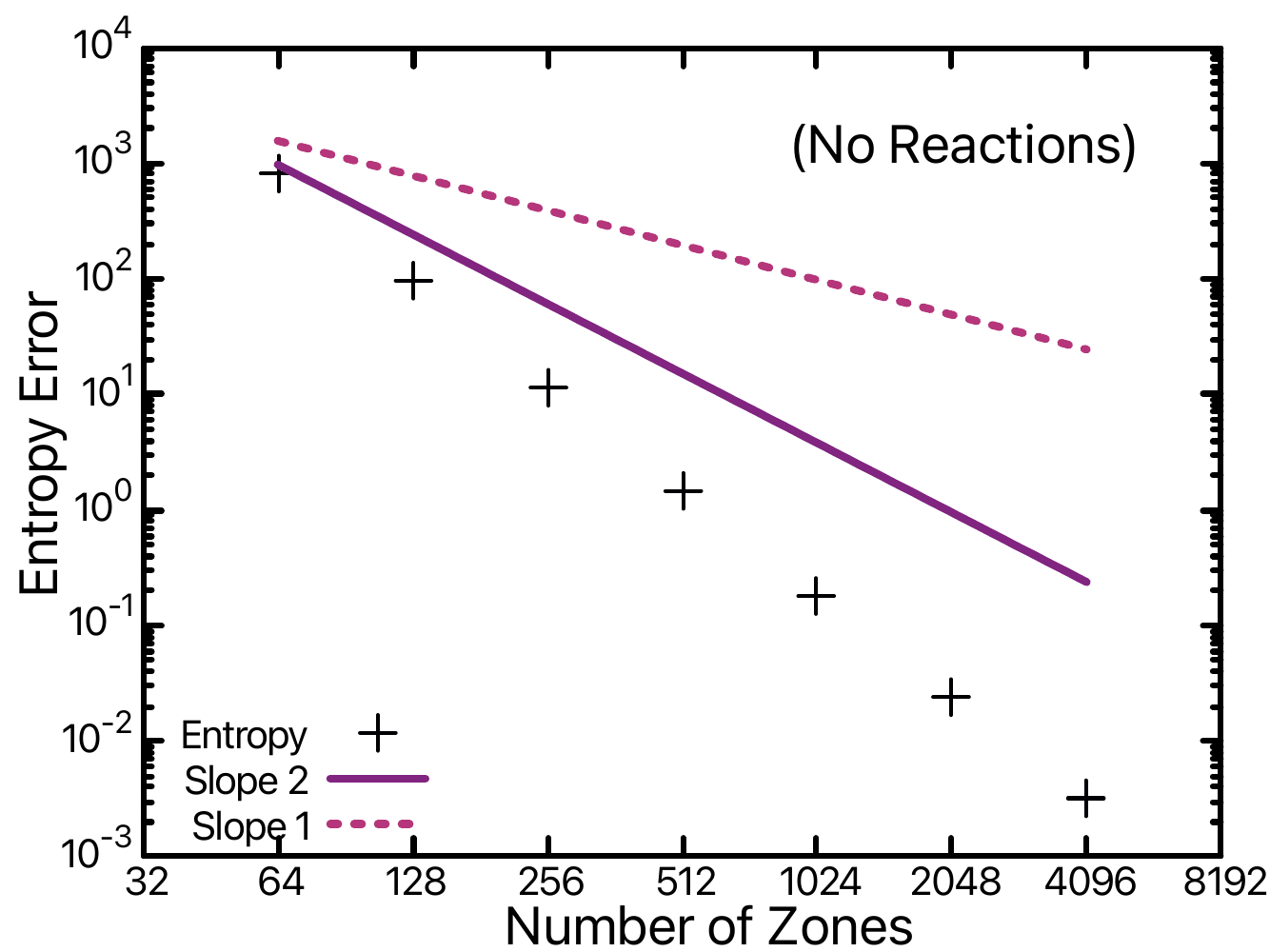}{0.4\textwidth}{}
\caption{$L_1$ entropy error vs. number of zones for the acoustic pulse test without burning. Convergence slopes of 1 and 2 are shown for comparison.}
\label{fig:AP_nore}
\end{figure}

To test the overall convergence of the testbed code, we run the same initial conditions over a variety of resolutions both with and without reactions. Without reactions, entropy should be conserved, and so the numerical entropy error for each resolution is calculated as a volume weighted $L_1$ norm over deviances with respect to the initial, ambient value. Equation \ref{eq:L1_nore} shows how the error is evaluated, and Figure \ref{fig:AP_nore} plots the error versus resolution. Note, we choose to plot resolution in terms of number of zones as the moving mesh technique results in individual zones growing and shrinking over the computation.

\begin{equation}
\label{eq:L1_nore}
    \begin{gathered}
        L_1(N) = \frac{\sum_{i=0}^N (\Delta x[i] \cdot |s_N[i]-s_0|)}{N \sum_{i=0}^N \Delta x[i]}, \\
        s_0 = \SI{3.484316488e8}{erg\per\gram\per\K}.
    \end{gathered}
\end{equation}

\begin{figure}
\centering
\fig{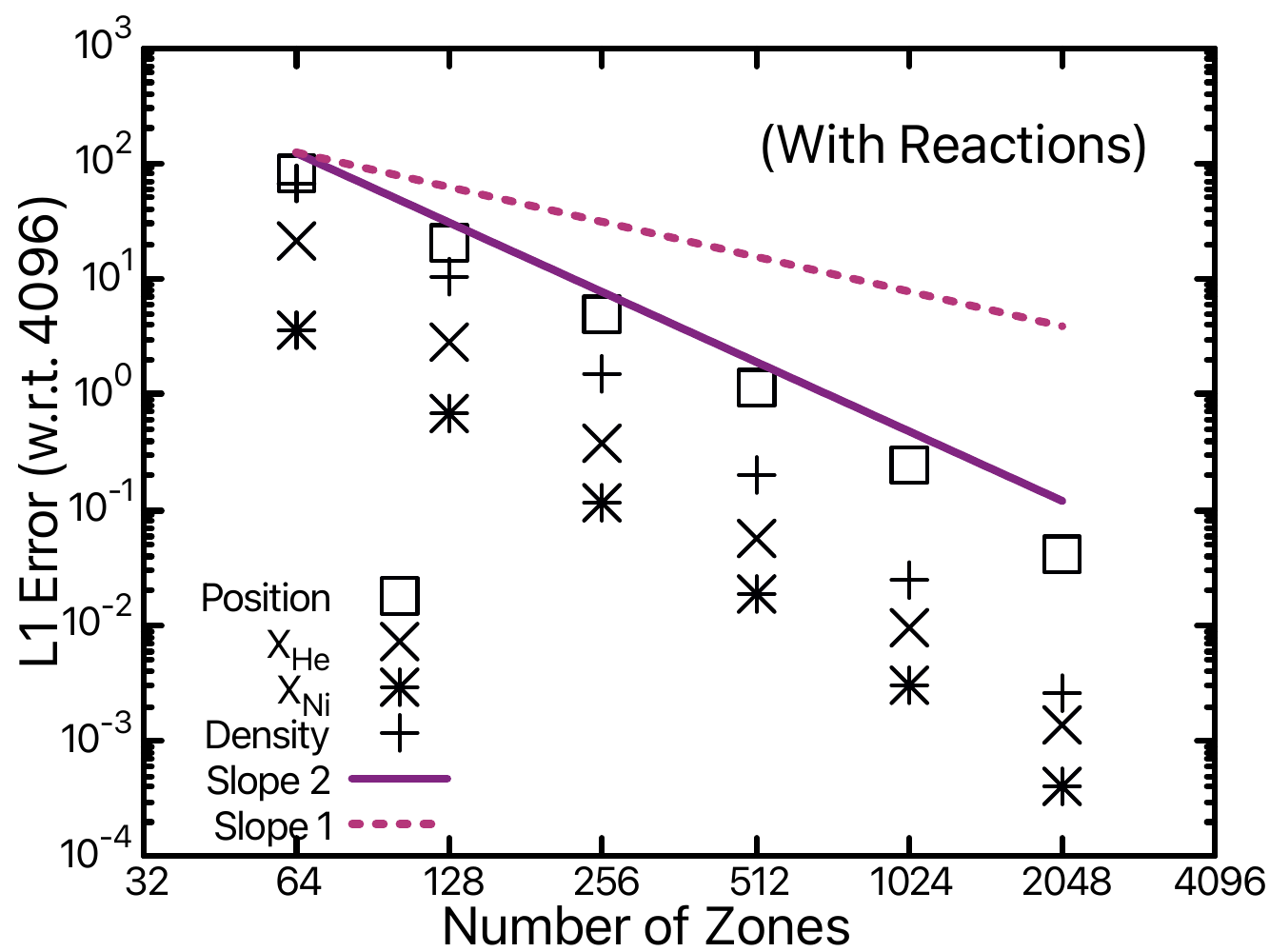}{0.4\textwidth}{}
\caption{Self-convergence in the acoustic pulse test with burning. $L_1$ error with respect to the 4096 run plotted over number of zones. The curves for position, $X$(He), and $X$(Ni) are shifted vertically for purposes of concise plotting by factors of $10^{-2.5}$, $10^{6.5}$, and $10^{39}$ respectively.}
\label{fig:AP_reac}
\end{figure}

In the case of hydrodynamics coupled with reactions, entropy is no longer conserved because the reactions release energy into the system; therefore, we switch to evaluating self-convergence. The highest resolution run (corresponding to 4096 zones) is used as the baseline when computing the volume weighted $L_1$ norms. Equation \ref{eq:L1_reac} shows how self-convergence error is evaluated for density, and in Figure \ref{fig:AP_reac}, we show a self-convergence plot for four different quantities: zone position, total density, mass fraction of helium, and mass fraction of nickel. The zone position is included among the quantities of interest because the moving mesh is not perfectly aligned with the highest resolution run. Instead, the zone positions will also converge at second order.

\begin{equation}
\label{eq:L1_reac}
    L_1(N) = \frac{\sum_{i=0}^N (\Delta x[i] \cdot \left|\rho_N[i]-\rho_{4096}[\frac{4096}{N}\cdot i]\right|)}{N\sum_{i=0}^N \Delta x[i]}.
\end{equation}

In addition to the convergence plots, Appendix \ref{sec:ap_tables} contains a table with the calculated errors and local rates for both the non-reacting and reacting cases. Pure hydrodynamics with the \texttt{helmeos} equation of state demonstrates second order convergence over the isentropic pulse. With reactions included, the self-convergence rates are also second order. As the computational methods are only second order, the rates above second will eventually taper off to a slope of 2.

\begin{figure*}
\centering
\fig{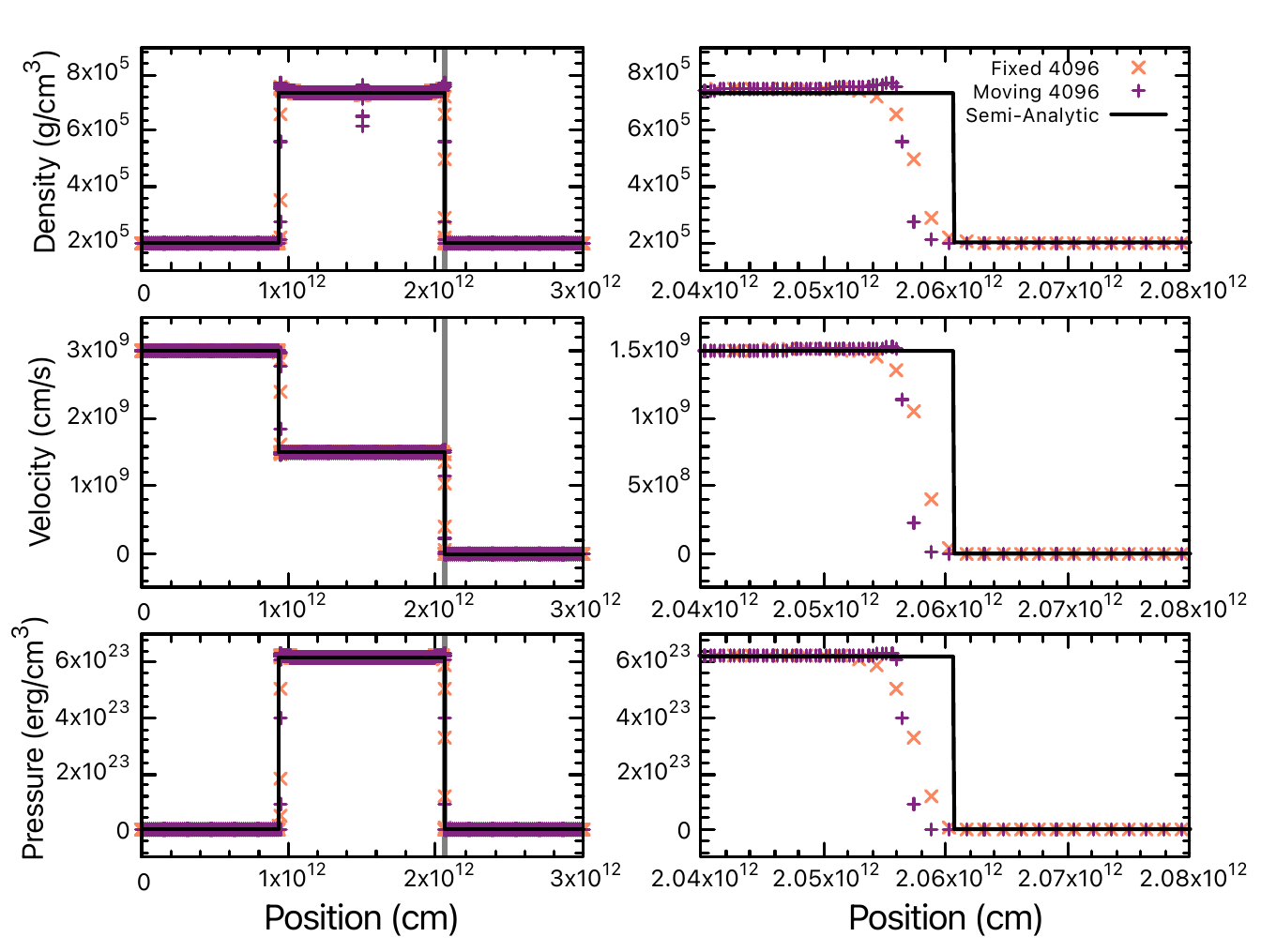}{0.75\textwidth}{}
\caption{Density, velocity, and pressure plotted over the domain for the burning shock problem for moving mesh, fixed mesh, and the semi-analytic solution. (\textit{Left}) The full domain shows the top-hat solution is achieved at 4096 zones. The gray column shows the zoomed region. (\textit{Right}) The region close to the right-most shock shows that both methods give shock velocities within $1\%$ of the semi-analytic solution.}
\label{fig:ShockFront}
\end{figure*}

\begin{figure}
    \centering
    \includegraphics[width=0.45\textwidth]{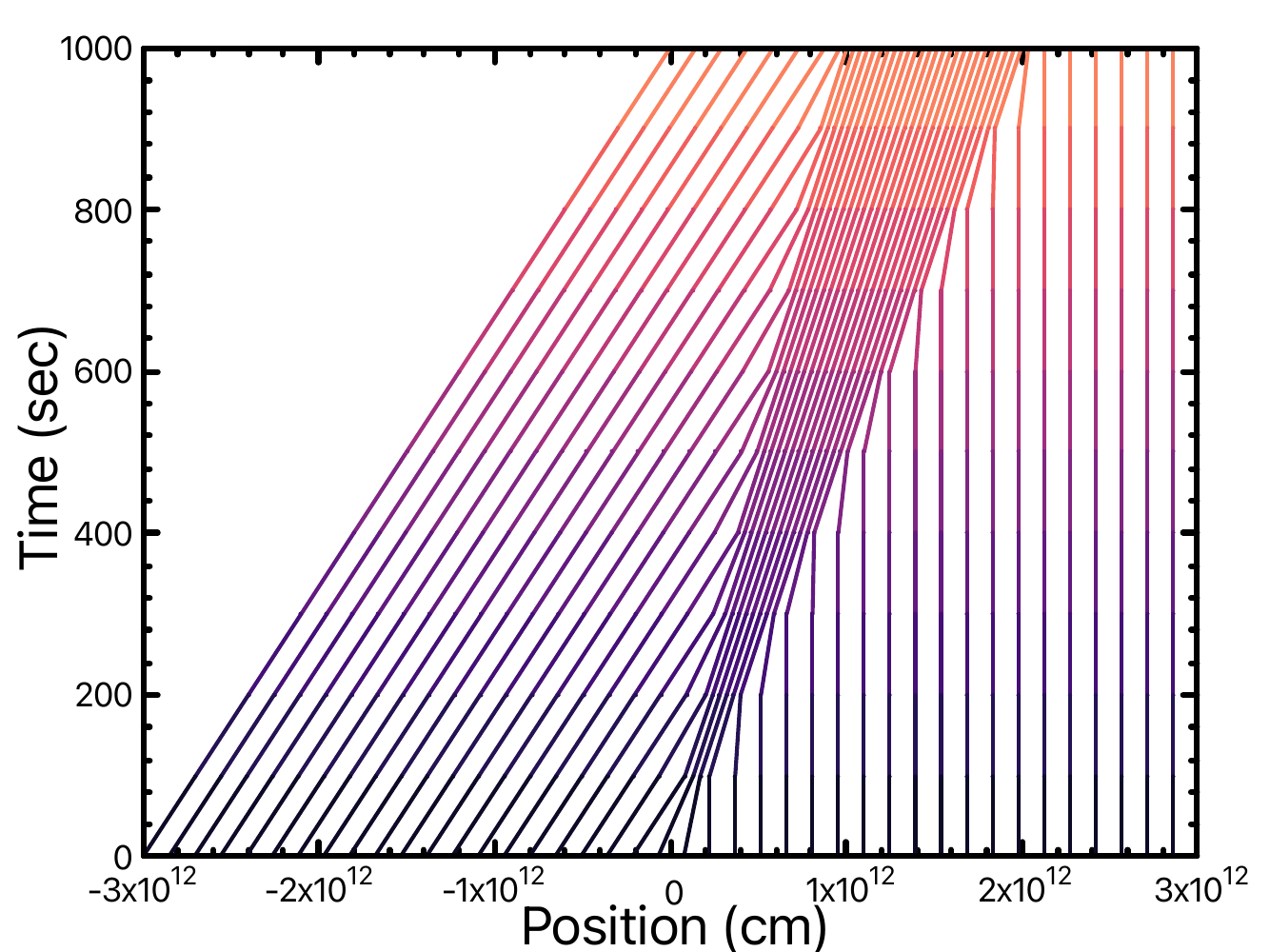}
    \caption{The burning shock mesh motion over time. Each line represents the trajectory of a zone. Only every 100th zone is plotted. The shocked region is compressed by a factor of $\sim 3.68$.}
    \label{fig:bs_mesh}
\end{figure}

\subsection{Burning Shock}
Here we present the numerical application of the test problem described in Section \ref{sec:detonation}. The collision, resulting from fast \SI{3e9}{\cm\per\sec} material slamming into the stationary material, is to be run for $t=\SI{1000}{\sec}$. The starting position of the velocity jump ($x_j$) of the cold material is centrally located in the domain $x=[-\SI{3e12}{\cm},\SI{3e12}{\cm}]$. Listed in Table \ref{table:burning_shock}, these initial conditions are chosen such that sufficient temperature is reached to burn $^4$He quickly to $^{56}$Ni behind the shock. 

\begin{table}[h!]
\centering
\begin{tabular}{||c c||}
 \hline
 Parameter & Value \\ [0.5ex] 
 \hline\hline
 $\rho_0$ & \SI{2e5}{\gram/\cm^3} \\ 
 $T_0$ & \SI{1e6}{\kelvin} \\
 $X_0(\mbox{He})$ & 1.0 \\
 $v_L$ & \SI{3e9}{\cm\per\sec}\\
 $v_R$ & \SI{0}{\cm\per\sec} \\
 $X_1(\mbox{Ni})$ & 1.0 \\
 $v_1$ & \SI{1.5e9}{\cm\per\sec}\\
 \hline
 $L$ & \SI{6e12}{\cm} \\
 $x_j$ & \SI{0}{\cm} \\
 $t$ & \SI{1000}{\sec} \\
 \hline
 $\rho_1$ & \SI{7.351e5}{\gram\per\cm^3} \\
 $T_1$ & \SI{3.293e9}{\kelvin} \\
 $v_{sh}$ & \SI{2.061e9}{\cm\per\sec} \\
 \hline
\end{tabular}
\caption{(\textit{Top}) Initial parameters of the burning shock problem. (\textit{Middle}) Domain parameters. (\textit{Bottom}) Semi-analytic solution values taking $v_1=\SI{1.5e9}{\cm\per\sec}$.}
\label{table:burning_shock}
\end{table}

In practice, the burning is not instantaneous, but takes $\sim 1-10 ~\unit{\sec}$ to reach $95\%$ nickel. However, both the fixed and moving mesh numerical computations, with adequate resolution, achieve top-hat like solutions. Figure \ref{fig:ShockFront} shows the $\rho$, $v$, and $P$, evaluated compared to the semi-analytic solution both over the entire domain and zoomed in on the right-most shock. 

Additionally, Figure \ref{fig:bs_mesh} shows the compression of the mesh. Across the shock, the density of zones is compressed by the same factor as the density of the material, in this case $\sim 3.68$. However, since we place no arbitrary lower bound on cell volume, the initial cold collision of supersonic material at the beginning of the run compresses two zones so that the volume is $1/10$ the initial volume instead of the expected $1/3.68$. These two zones then dominate the timestep of the moving mesh case and result in the moving mesh case taking $\sim 41\%$ longer than the fixed mesh. The total CPU time in the 512-zone run for the moving mesh is $9.27 \times 10^5$ seconds while the fixed mesh is $6.55 \times 10^5$ seconds.

\begin{figure}
\centering
\fig{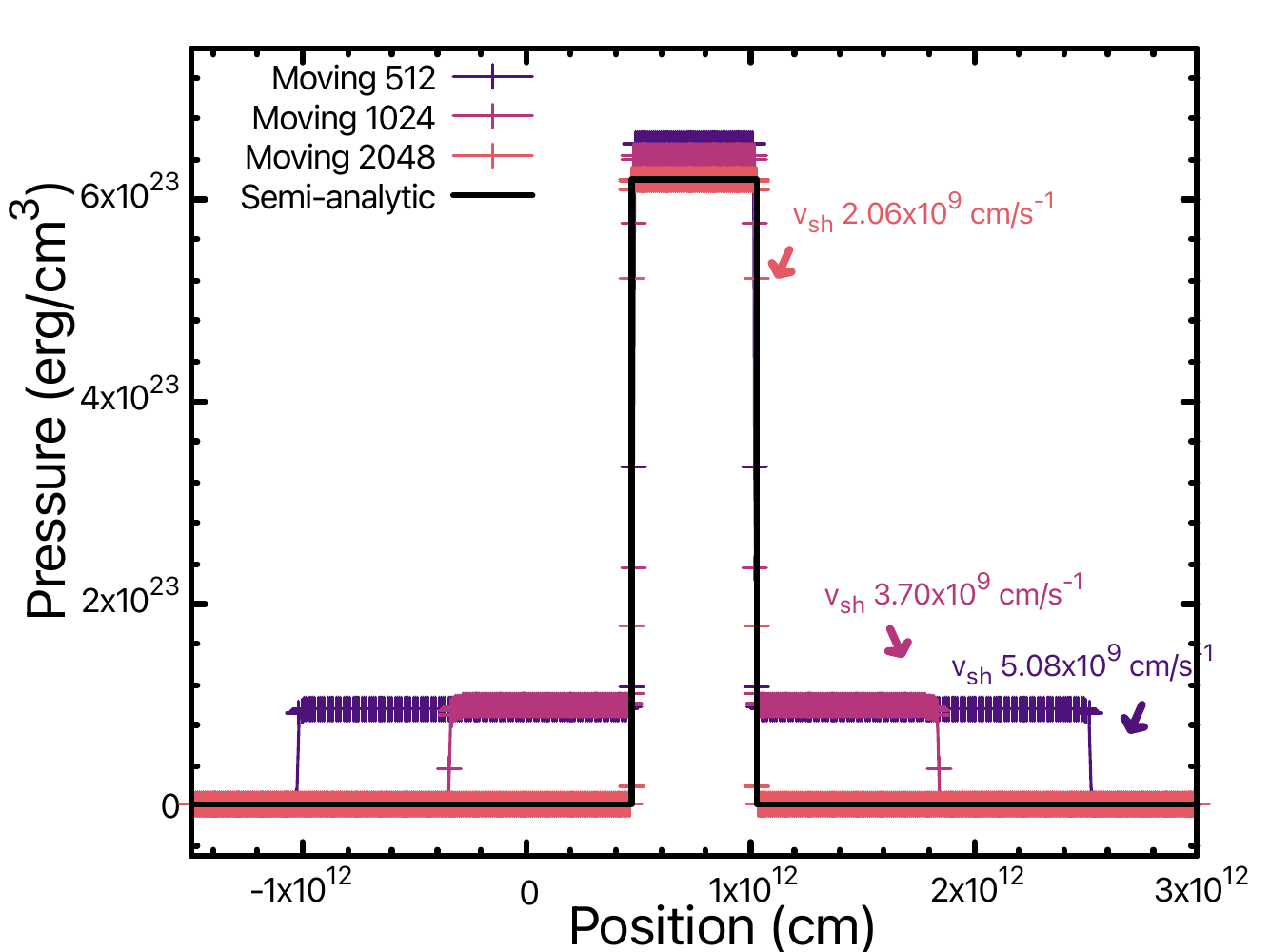}{.4\textwidth}{}
\caption{Checkpoint at $t=\SI{2e-4}{\sec}$. The moving mesh burning shock pressure top-hat is resolved by the 2048 run. The spurious shock speed is labeled by its corresponding self-consistent value for the unresolved cases.}
\label{fig:ShockFrontDegenMov}
\end{figure}

\begin{figure}
\centering
\fig{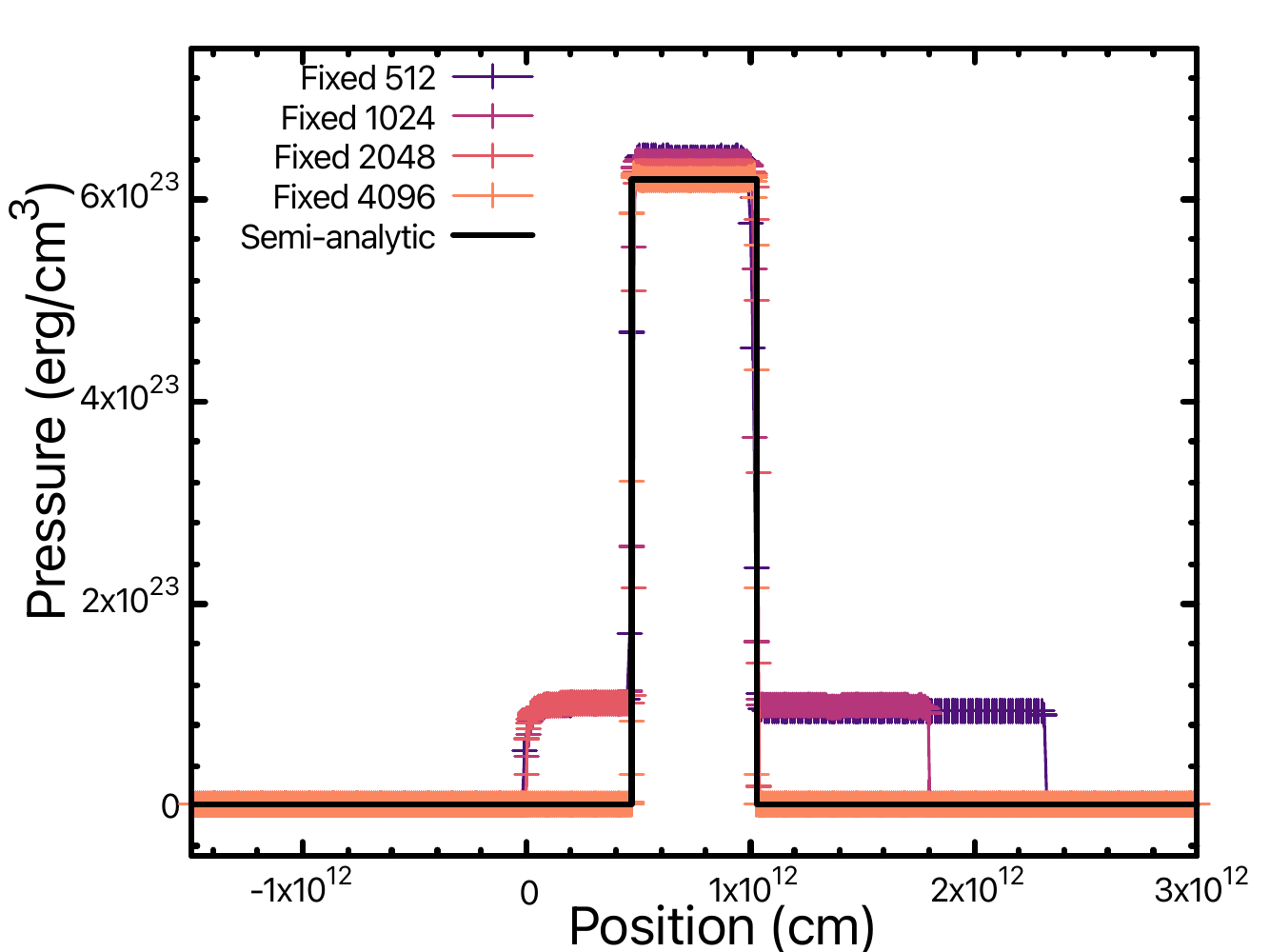}{.4\textwidth}{}
\caption{Checkpoint at $t=\SI{2e-4}{\sec}$. The fixed mesh burning shock pressure top-hat is resolved by the 4096 run. The spurious shocks are seen in the 512 and 1024 cases, while the 2048 case shows only a lingering artifact beginning at $x=0$.}
\label{fig:ShockFrontDegenFix}
\end{figure}

At low resolutions, spurious, ultra-fast shocks proceed ahead of the expected shock. These spurious shocks do not disappear by decreasing the $C_{\mbox{CFL}}$ number (i.e. shortening the timestepping and improving the reaction-hydrodynamics coupling), but they do vanish with increasing resolution. This is seen in Figure \ref{fig:ShockFrontDegenMov} and Figure \ref{fig:ShockFrontDegenFix}. Each jump visible is consistent with jump conditions in Section \ref{sec:detonation}, but with a different velocity behind the shock and resulting composition. Upon closer inspection, moving mesh first achieves the single top-hat solution starting with $\sim 1722$ zones, while fixed mesh requires $\sim 2460$ or more zones. We attribute this benefit to the natural compression of zones in the shocked region that the moving mesh technique provides for this test problem.  

Note the narrow, low density and high temperature feature at the center of the density top hat (top left panel) of Figure \ref{fig:ShockFront}. This is a numerical artifact that diminishes with increasing resolution and results from the initial unresolvable velocity discontinuity. It has been investigated in a similar 1D problem as a source of early ignition in white dwarf head on collisions (\cite{Kushnir_2013}, \cite{Katz_2019}, and \cite{Kushnir_2019}). Considering our initial conditions, the hydrodynamical shock alone results in a temperature of $\sim \SI{3e9}{K}$, so the shocked region will burn quickly and completely regardless of the numerical artifact. However, in the early time formation of the shocked region the high temperature nature of the artifact could inflate the burning rates. We ran a nonuniform initial distribution of zones and found that the spurious, fast shocks still form far away from this artifact, so this early time, high temperature is at least not the sole reason for the ultra-fast shocks. A certain resolution is required for steady propagation of the detonation. 

\subsection{1D Deflagration \& Advection}

The final test problem is motivated by the proof-of-concept in \cite{Zingale_2019}. The problem is a smooth transition from fuel to ash where the mixed region eventually ignites and, through the process of thermal diffusion, drives a deflagration flame front. In this work, we modify the original problem by mirroring the domain in order to apply a periodic boundary condition. Then we introduce a global bulk velocity that advects the domain a predetermined amount. Faster bulk velocity will result in more numerical diffusion, which in turn can drive artificial deflagration. 

The problem is initialized by picking a fuel state $\rho_0$, $T_0$, and $X_0(\mbox{He})$. The equation of state is used find the ambient pressure:

\begin{equation}   
    P_0=P^{(eos)}(\rho_0,T_0,\mathbf{X}_0).
\end{equation}

Next, we assign the ash state temperature and composition to be $T_1$ and $X_1(\mbox{Ni})$. We smoothly transition between the two states via a tanh function:

\begin{equation}
    \label{eq:def_tanh}
    \begin{gathered}
        T(x) = T_0 + \frac{1}{2}\left( T_1-T_0\right) \left[ 1 - \tanh \left(\frac{x-x_c}{\delta}\right) \right] \\
        \mathbf{X}(x) = \mathbf{X}_0 + \frac{1}{2}\left( \mathbf{X}_1-\mathbf{X}_0\right) \left[ 1 - \tanh \left(\frac{x-x_c}{\delta}\right) \right].
    \end{gathered}
\end{equation}

Finally using this temperature, composition, and the ambient pressure, the equation of state sets the density and the remaining hydrodynamic quantities:

\begin{equation}
    \rho(x) = \rho^{(eos)}\left(P_0,T(x),\mathbf{X}(x)\right).
\end{equation}

The two free parameters $x_c$ and $\delta$ that appear in Equation \ref{eq:def_tanh} control the location and width of the transition from ash to fuel. They are set in reference to $L_x$, which is half the total domain. Lastly, we mirror about $x = L_x$ to fill out the second half of the domain. 

\begin{table}[h!]
\centering
\begin{tabular}{||c c||}
 \hline
 Parameter & Value \\ [0.5ex] 
 \hline\hline
 $\rho_0$ & \SI{2e7}{\gram/\cm^3} \\ 
 $T_0$ & \SI{5e7}{\kelvin} \\
 $X_0(\mbox{He})$ & 1.0 \\
 $v_{\mbox{bulk}}$ & \SI{6.4e6}{} OR 0.0 \unit{\cm/\sec}\\
 $T_1$ & \SI{3.6e9}{\kelvin} \\
 $X_1(\mbox{Ni})$ & 1.0 \\
 \hline
 $x_c$ & $0.4L_x$ \\
 $\delta$ & $0.06L_x$ \\
 $L_x$ & \SI{256}{\cm} \\
 $t$ & \SI{4e-4}{\sec} \\
 \hline
\end{tabular}
\caption{(\textit{Top}) Initial parameters of the deflagration problem. (\textit{Bottom}) Domain parameters.}
\label{table:deflagration}
\end{table}

\begin{figure*}
\centering
\fig{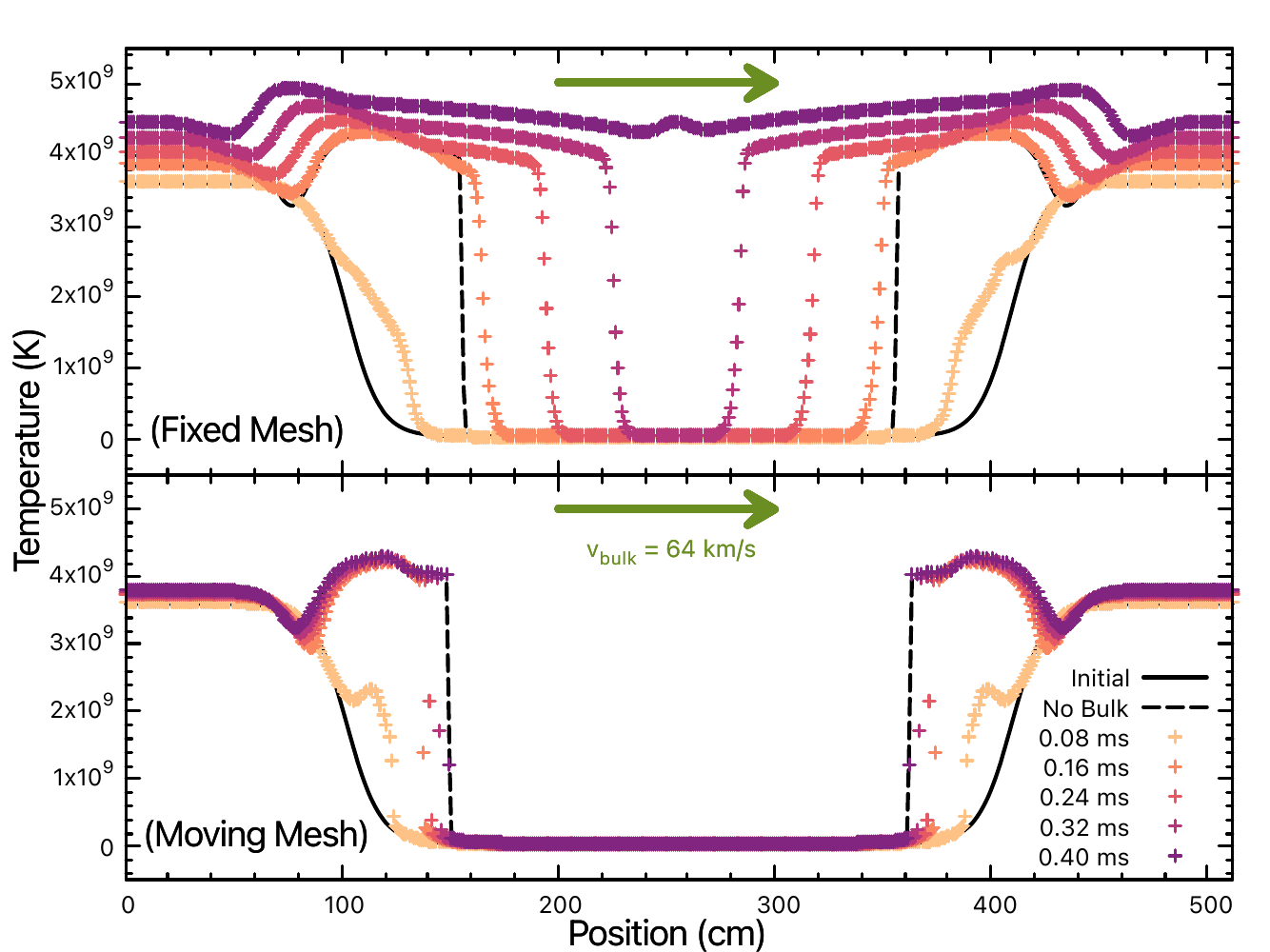}{.75\textwidth}{}
\caption{The advecting-deflagration flame front test: temperature curves for a bulk velocity of \SI{6.4}{\km\per\sec} at a resolution of 512 zones are plotted with 5 checkpoints. The dashed lines are the end results of corresponding zero bulk velocity runs. (\textit{Top}) Fixed mesh results in an inflated propagation rate of \SI{3.95}{\km\per\sec} as the thermal diffusion is vastly subdominant to the numerical diffusion. (\textit{Bottom}) Moving mesh results in a propagation speed of \SI{0.525}{\km\per\sec} both with and without advection.}
\label{fig:Deflagration}
\end{figure*}

To compare fixed vs moving mesh, we ran this setup above with a bulk velocity of $v_{\mbox{bulk}} = \SI{6.4e6}{\cm\per\sec}$ for a total time $t = \SI{4e-4}{\sec}$. This corresponds to the entire fluid cycling fives times through the domain. Figure \ref{fig:Deflagration} displays the results. The region of primary interest is the deflagration flame front marked by the rapid transition of cold fuel to fresh and hot ash at $\sim \SI{4E9}{\K}$ that propagates towards the center of the domain over time. 

In Figure \ref{fig:Deflagration}, we see the artificially enhanced deflagration of the fixed mesh propagate throughout the entire fuel region. Meanwhile, the moving mesh effectively negates the numerical diffusion from the bulk advection. As a reference, the final checkpoint of each respective zero bulk velocity run is included as a dashed line, which emphasizes how the fixed mesh numerical diffusion inflates the deflagration front speed. 

\begin{figure}
    \centering
    \includegraphics[width=0.45\textwidth]{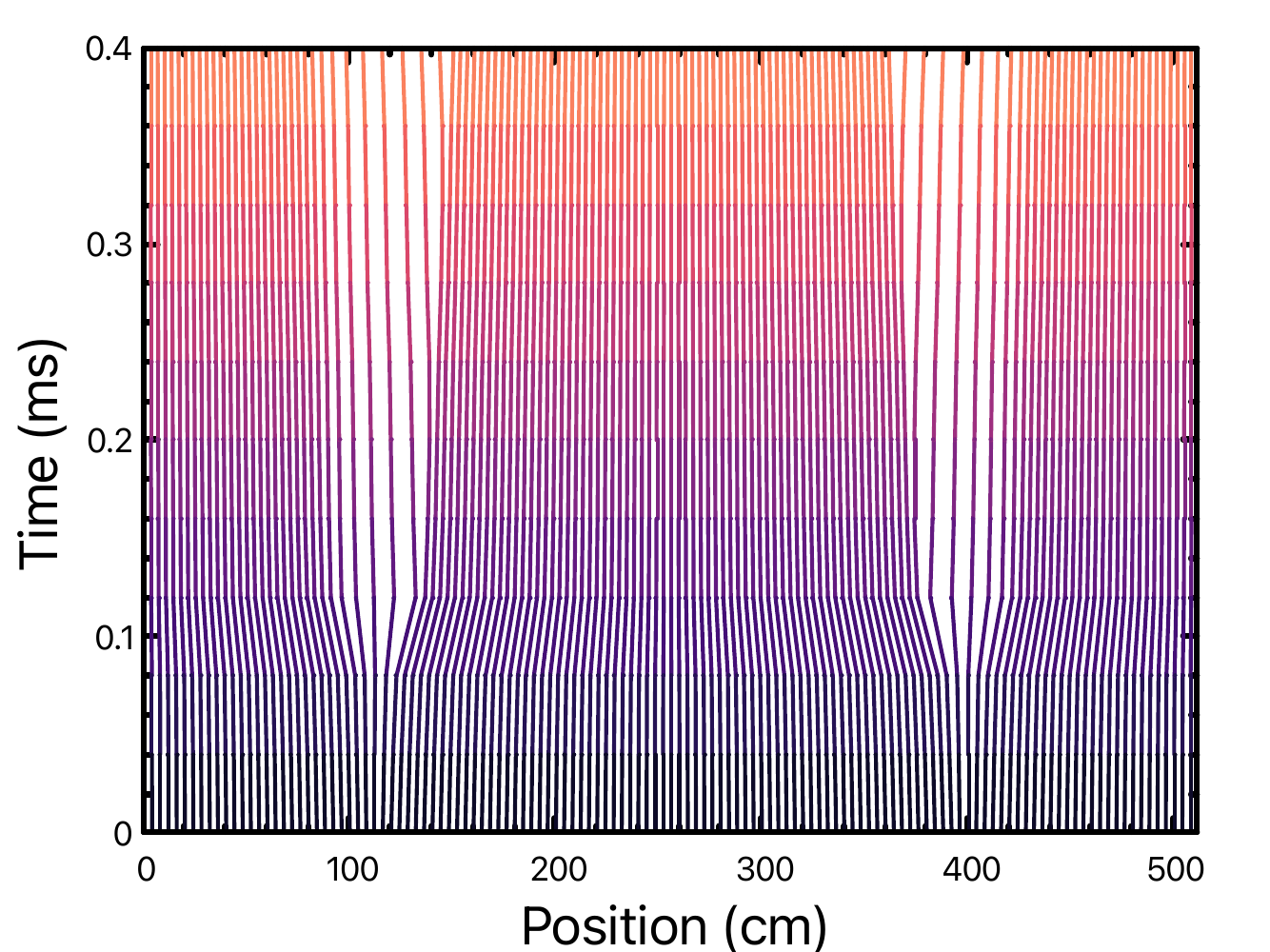}
    \caption{The deflagration mesh motion over time in the no bulk velocity case. Each line represents the trajectory of a zone. Only every fourth zone is plotted.}
    \label{fig:def_mesh}
\end{figure}

Figure \ref{fig:def_mesh} shows the no bulk velocity mesh motion. There is expansion in the two active burning regions as the energy from reactions is released and compression elsewhere. The minimum volume reached is $0.845 \times$ the initial volume. Once again however, the subtraction of the average mesh motion allows the moving mesh to take a net longer timestep. In the 256-zone run, the total CPU time for the moving mesh is $1.57 \times 10^7$ seconds and $1.70 \times 10^7$ seconds for the fixed mesh.

\begin{figure}
    \fig{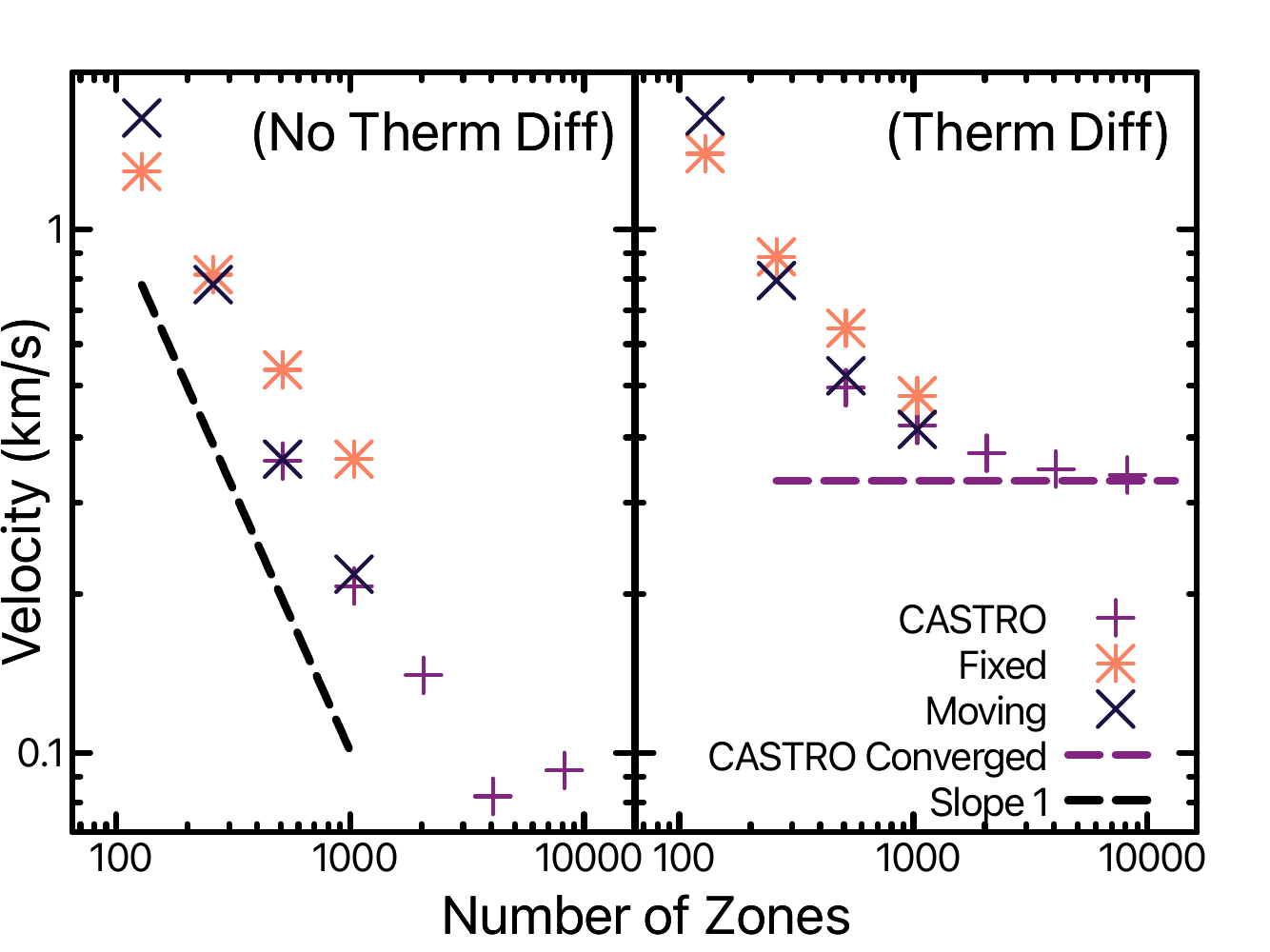}{0.4\textwidth}{}
    \caption{Deflagration speeds with resolution, for the deflagration test with no bulk velocity. The moving mesh method appears to converge at a slightly higher rate than fixed mesh and shows dramatic improvement in runs beyond 256 zones. (\textit{Left}) No thermal diffusion is included. By 1024 zones the numerical diffusion is subdominant to the thermal diffusion. The final three \texttt{CASTRO} points are determined only with the final 2 checkpoints as the flame front only forms right at the end. (\textit{Right}) Thermal diffusion is included into the energy equation. \texttt{CASTRO} converges to $\SI{3.3e4}{\cm\per\sec}$ at first order.}
    \label{fig:def_speed}
\end{figure}

Comparing the zero bulk velocity dashed lines, we observe that the moving mesh technique has a slightly slower deflagration speed at this resolution. The released energy causes some expansion in the new ash region, and this induced velocity introduces numerical diffusion that can artificially inflate the deflagration speed even in the bulk static case. We run our code both with and without thermal diffusion to explore the underlining artificial deflagration versus real deflagration and quantify any further moving mesh advantage. As this becomes computationally expensive for this testbed code, we also run this test on the nuclear hydrodynamics code \texttt{CASTRO} to confirm that the speed of the deflagration flame front eventually converges. \texttt{CASTRO} is run in 1D, in the Strang$+$CTU integration mode, without AMR, with a $C_{\mbox{CFL}}$ number of $0.2$, without bulk advection, and up to a much higher resolution of $8192$ zones. Our fixed mesh case still differs from \texttt{CASTRO} in part because \texttt{CASTRO} uses a distinct (yet similar to \texttt{aprox13}) 13 isotope reaction network, and \texttt{CASTRO} implements thermal diffusion via a source term construction. The speed of the flame front is determined by tracking the position of the final six checkpoints. We linearly extrapolate to find the position where temperature crosses $\SI{2e9}{K}$. Then we take the average of the five velocities resulting from these six checkpoints as the speed for any given run. Figure \ref{fig:def_speed} displays the many computations both with and without thermal diffusion. 

\texttt{CASTRO} converges to velocity of the deflagration of $\SI{3.3e4}{\cm\per\sec}$ at first order, and our testbed code is consistent with this result. Comparing our fixed mesh to moving mesh, we do see a dramatic improvement in reduction of numerical diffusion at middling resolutions as it appears moving mesh is converging to a velocity at a steeper rate. Note at the higher resolutions in Figure \ref{fig:def_speed}, the speed of the artificial deflagration with no thermal diffusion is significantly below the value obtained when including thermal diffusion, which implies our testbed code is well on its way to convergence. 

\section{Discussion} \label{sec:discussion}
In this work, we present applications of the moving mesh technique to nuclear hydrodynamics via our 1D Cartesian testbed code coupled to a 13 isotope network. The aim is quantify the advantages of the moving mesh by direct comparisons to the fixed mesh case. For completeness and availability, a ``frozen" version of this code will be made available on Zenodo.

First with the acoustic pulse test problem, we demonstrate that our numerical methodology does indeed achieve second order convergence in entropy conservation in the pure hydrodynamics + stellar equation of state case. With nuclear reactions integrated along with hydrodynamics, our testbed code shows a self-convergence rate $\sim 1.8$ in density and abundances using traditional Strang splitting. 

In Section \ref{sec:detonation}, we present an semi-analytic solution to a burning front test problem in the limit of quick burning. As quick burning itself is a difficult problem, this is an excellent test problem to explore the impact of resolution, integration methods, burning limiters, etc. in the future. In this work, we see the moving mesh technique does provide an advantage in resolution because of the natural compression of the region of interest. 

With the addition of thermal diffusion, we explore how moving mesh can reduce numerical diffusion and the artificial mixing of fuel and ash in the deflagration problem. In the case of bulk advection, the moving mesh technique dramatically reduces the artificial mixing and preserves the stationary solution. Upon closer inspection of the static deflagration case, moving mesh does provide advantages in reducing the numerical diffusion at middling to high resolutions. 

This hints at future applications of the moving mesh technique providing extraordinary improvement in large scale astrophysical flows or cases with two bodies that necessitates motion with respect to a fixed mesh. The immediate future step is to incorporate this nuclear reaction methodology into the spherically symmetric code \texttt{RT1D} \citep{Duffell_2016b} to begin modeling type Ia supernovae. A multidimensional extension of this method has already been demonstrated using a dynamic Voronoi tessellation in the \texttt{AREPO} code \citep{Pakmor_2022}. \texttt{AREPO} has been exploring moving mesh nuclear hydrodynamics and has computed a violent merger scenario for detonating white dwarves \citep{Pakmor_2013}. We intend to extend our work to multidimensions using the more structured grids of the \texttt{JET} \citep{Duffell_2018} and \texttt{SPROUT} \citep{Mandal_2023} codes. Both have meshes designed to alleviate computational burden while retaining the benefits of moving mesh in specific scenarios, like radial flows in spherical coordinates for \texttt{JET} and homologous expansions in Cartesian coordinates for \texttt{SPROUT}.

\begin{acknowledgments}
We thank Michael Zingale and Dean Townsley for their constructive comments on the test problems. We also thank Abigail Polin for advice and assistance with \texttt{CASTRO} calculations. Further, we thank Danielle Dickinson for helpful comments on the manuscript. Numerical calculations were preformed on the Petunia computing cluster hosted by the Department of Physics and Astronomy at Purdue University. We thank Chris J. Orr for his extensive help with maintaining Petunia. 
\end{acknowledgments}

%

\vspace{5mm}





\appendix

\section{Fourth Order Semi-Implicit Integrator} \label{sec:integrator}

Within the context of Strang Splitting to couple hydrodynamics with nuclear reaction networks, we need to evolve a given composition of isotopes $\mathbf{Y}$ at some density $\rho$ and temperature $T$, hydrostatically (i.e. $\rho$ and $T$ are constant over the integration). The molar abundances $\mathbf{Y}$ are obtained from the mass fractions $\mathbf{X}$ by $Y_i = X_i/A_i$ where $A_i$ is the mass number of isotope $i$. Since reaction networks are generally stiff ODEs, it is beneficial to use a high order implicit method. Our basic step follows the semi-implicit Backward Euler formulation to advance a composition by time $h$.
\begin{equation} 
\mathbf{Y}_f=\mathbf{Y}_i + h \left[ 1-h\frac{\partial f}{\partial Y} \Big|_{\mathbf{Y}_i} \right]^{-1}\cdot \mathbf{f}(\mathbf{Y}_i)
\label{eq:semi_impli_step}
\end{equation}
Where $\mathbf{Y}_f$ is the final composition, $\mathbf{Y}_i$ is the initial composition, $\frac{\partial f}{\partial Y}$ is the corresponding Jacobian of the ODE, and $\mathbf{f}(\mathbf{Y}_i)$ is the corresponding right hand side of the ODE. To construct a $4^{th}$ order method, we break the full interval into smaller, equidistant sub-steps. We can now construct a system of equations using the $2^{nd}$, $3^{rd}$, and $4^{th}$ order error terms taking one full step, two half steps, three third steps, and four fourth steps. 
\begin{equation} 
\begin{array}{llll}
\mathbf{Y}(x+4h) = \mathbf{Y}_1+(4h)^2\theta+(4h)^3\phi+(4h)^4\psi+O(h^5) \\
\mathbf{Y}(x+4h) = \mathbf{Y}_2+2(2h)^2\theta+2(2h)^3\phi+2(2h)^4\psi+O(h^5)\\
\mathbf{Y}(x+4h) = \mathbf{Y}_3+3(\frac{4}{3}h)^2\theta+3(\frac{4}{3}h)^3\phi+3(\frac{4}{3}h)^4\psi+O(h^5)\\
\mathbf{Y}(x+4h) = \mathbf{Y}_4+4h^2\theta+4h^3\phi+4h^4\psi+O(h^5)
\end{array}
\end{equation}
Where $\mathbf{Y}(x+4h)$ is the true solution after advancing time by $4h$, $\mathbf{Y}_i$ is the result of taking $i$ consecutive steps of time $\frac{4}{i}h$, and $\theta$, $\phi$, $\psi$ are the $2^{nd}$, $3^{rd}$, and $4^{th}$ error terms, respectively. Note that this method requires 7 evaluations of the Jacobian and right hand side and 10 evaluations of Equation \ref{eq:semi_impli_step} in total. From here, it is simple to solve the four equations and four unknowns to find a solution of the following form.
\begin{equation} 
\begin{array}{ccc}
\mathbf{Y}(x+4h) = \mathbf{Y}_{3^{rd}}+\Delta_{4^{th}}+O(h^5) \\
\mathbf{Y}_{3^{rd}} = 8\mathbf{Y}_4-9\mathbf{Y}_3+2\mathbf{Y}_2 \sim h^3\\
\Delta_{4^{th}} = \frac{8}{3}\mathbf{Y}_4-\frac{9}{2}\mathbf{Y}_3+2\mathbf{Y}_2-\frac{1}{6}\mathbf{Y}_1 \sim h^4
\end{array}
\end{equation}
Where $\mathbf{Y}_{3^{rd}}$ is a third order estimate of the solution, $\Delta_{4^{th}}$ is the local truncation error, and $(\mathbf{Y}_{3^{rd}}+\Delta_{4^{th}})$ will be our fourth order solution. This formulation is useful to implement adaptive stepping. Consider integrating the entire network over a total time of $\Delta t$ with the desire to keep error beneath some globally set relative and absolute tolerances $RTOL$ and $ATOL$. Now we can attempt an integration using the method described above for some trial timestep $h_1$ and calculate a new timestep using the resulting truncation error and our desired errors. 
\begin{equation} 
\begin{array}{cc}
h_0 = h_1\Big| \frac{\Delta_0}{\Delta_{4^{th}}} \Big|^{1/4},\\
\Delta_0 = RTOL\cdot |\mathbf{Y}_i|+ATOL
\end{array}
\end{equation}
Where $h_0$ is the newly scaled timestep, $\Delta_0$ is the desired accuracy scaled by the initial composition. If the truncation error is below the desired tolerances, then $h_0$ will increase and the step will be considered a success. We will update the composition to be $(\mathbf{Y}_{3^{rd}}+\Delta_{4^{th}})$ and progress time by $h_1$. The next step will integrate starting with the new composition and trial timestep. Else if the truncation error is greater than the desired tolerances, $h_0$ will decrease and the step will be rejected. The composition will not be updated, and the integrator will retry using the new smaller timestep. In practice, it is reasonable to limit the change in $h$ to a factor of two so that the integrator will not wildly adjust the trial timestep. Finally, the integrator proceeds until the composition has been updated to the final time ($t+\Delta t$).

\section{Acoustic Pulse Rate Tables} 
\label{sec:ap_tables}

This appendix includes the error and rate tables for the acoustic pulse convergence plots. The errors for the non-reacting and reacting cases are calculated by Equation \ref{eq:L1_nore} and Equation \ref{eq:L1_reac} respectively. The rates are evaluated as the logarithmic slope between two consecutive errors.
\begin{table}[h!]
\centering
\begin{tabular}{||c | c c||}
 \hline
 N & err$_{s}$ & Rate \\ [0.5ex] 
 \hline\hline
 64 & $8.041 \times 10^2$ &  \\ 
 128 & $9.516 \times 10^1$ & 3.079 \\
 256 & $1.161 \times 10^1$ & 3.035 \\
 512 & $1.444 \times 10^0$ & 3.008 \\
 1024 & $1.830 \times 10^{-1}$ & 2.980 \\
 2048 & $2.383 \times 10^{-2}$ & 2.941 \\
 4096 & $3.237 \times 10^{-3}$ & 2.880 \\
 \hline
\end{tabular}
\caption{The volume averaged entropy error for the non-reacting acoustic pulse. The rate is the local slope in logarithmic space of the respective row with the prior row.}
\label{table:ap_nore_conv}
\end{table}

\begin{table*}[h!]
\centering
\begin{tabular}{||c | c c | c c | c c | c c|| }
 \hline
 N & err$_{x}$ & Rate & err$_{\rho}$ & Rate & err$_{X_{He}}$ & Rate & err$_{X_{Ni}}$ & Rate\\ [0.5ex] 
 \hline\hline
 64 & $2.579 \times 10^4$ & & $6.663 \times 10^1$ & & $6.831 \times 10^{-6}$ & & $3.577 \times 10^{-39}$ & \\ 
 128 & $6.366 \times 10^3$ & 2.018 & $1.044 \times 10^1$ & 2.674 & $8.795 \times 10^{-7}$ & 2.957 & $6.860 \times 10^{-40}$ & 2.383 \\
 256 & $1.542 \times 10^3$ & 2.046 & $1.502 \times 10^0$ & 2.798 & $1.180 \times 10^{-7}$ & 2.897 & $1.150 \times 10^{-40}$ & 2.577 \\
 512 & $3.598 \times 10^2$ & 2.099 & $1.988 \times 10^{-1}$ & 2.917 & $1.825 \times 10^{-8}$ & 2.693 & $1.884 \times 10^{-41}$ & 2.610 \\
 1024 & $7.710 \times 10^1$ & 2.222 & $2.499 \times 10^{-2}$ & 2.992 & $3.033 \times 10^{-9}$ & 2.589 & $2.997 \times 10^{-42}$ & 2.652 \\
 2048 & $1.285 \times 10^1$ & 2.585 & $2.572 \times 10^{-3}$ & 3.280 & $4.349 \times 10^{-10}$ & 2.802 & $4.037 \times 10^{-43}$ & 2.892 \\
 \hline
\end{tabular}
\caption{The volume averaged error for reacting acoustic pulse problem over zone position, density, mass fraction of He, and mass fration of Ni. The error is calculated in reference to the $N=4096$ run. The rate is the local slope in logarithmic space of the respective row with the prior row.}
\label{table:ap_reac_conv}
\end{table*}

\bibliography{main}{}
\bibliographystyle{aasjournal}



\end{document}